\documentclass[9pt]{article}
\usepackage{mathptmx}
\usepackage{amsthm}
\usepackage{spconf,amsmath,graphicx}
\usepackage{enumitem}
\usepackage{epstopdf}
\usepackage{lipsum}
\usepackage{todonotes}
\usepackage{textcomp}
\usepackage{xspace}
\usepackage[perc]{overpic}
\usepackage{color}
\usepackage[absolute]{textpos}
\usepackage[bottom]{footmisc}
\usepackage{multicol}
\usepackage{multirow}
\usepackage[export]{adjustbox}
\usepackage{url}
\usepackage{arydshln}
\usetikzlibrary{spy}

\newcommand{\compactsubsub}[2]{\vspace{0.5mm}\noindent \textbf{\textit{#1:}} #2}

\newcommand{\chapterspacing}{-2mm}

\tikzset{annot/.style={draw=black,fill=white,text=black}}

\makeatletter
\DeclareRobustCommand\onedot{\futurelet\@let@token\@onedot}
\def\@onedot{\ifx\@let@token.\else.\null\fi\xspace}

\makeatother

\title{{Leveraging Self-supervised Denoising for Image Segmentation}}
\name{
Mangal Prakash$^{1,2}$, Tim-Oliver Buchholz$^{1,2}$, Manan Lalit$^{1,2}$, Pavel Tomancak$^{2}$,}
\secondlinename{Florian Jug$^{1,2\ast}$, Alexander Krull$^{1,2\ast}$}
\address{
$^{1}$Center for Systems Biology Dresden (CSBD)\\
$^{2}$Max-Planck Institute of Molecular Cell Biology and Genetics\\
$^{\ast}$joint supervision
}

\newcommand\figTeaser{
\begin{figure}[tb]
\centerline{
\includegraphics[width=\linewidth]{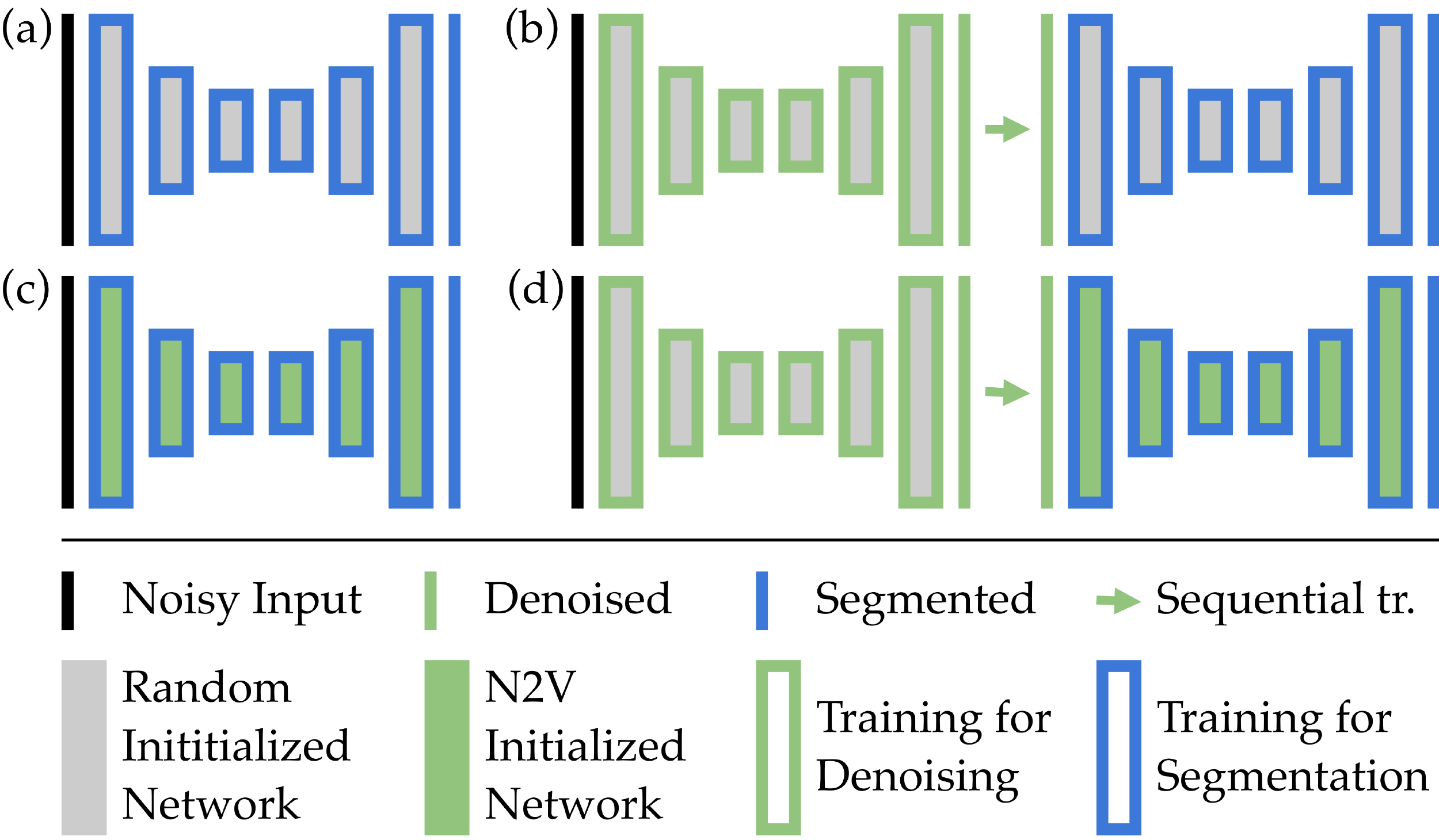}
}
\caption{Tested network architectures and training schedules. 
\textbf{(a)}~Baseline methods are directly trained to segment noisy data, 
\textbf{(b)}~sequential setup, with denoising being the preprocessing step for subsequent segmentation, 
\textbf{(c)}~finetuning of a pretrained denoising network for segmentation, and 
\textbf{(d)}~finetune-sequential, combining the ideas of (b) and (c).}
  \label{fig:Teaser}
\end{figure}
}

\newcommand\figDSBSEGAP{
\begin{figure}[t]
\centerline{
\includegraphics[width=\linewidth]{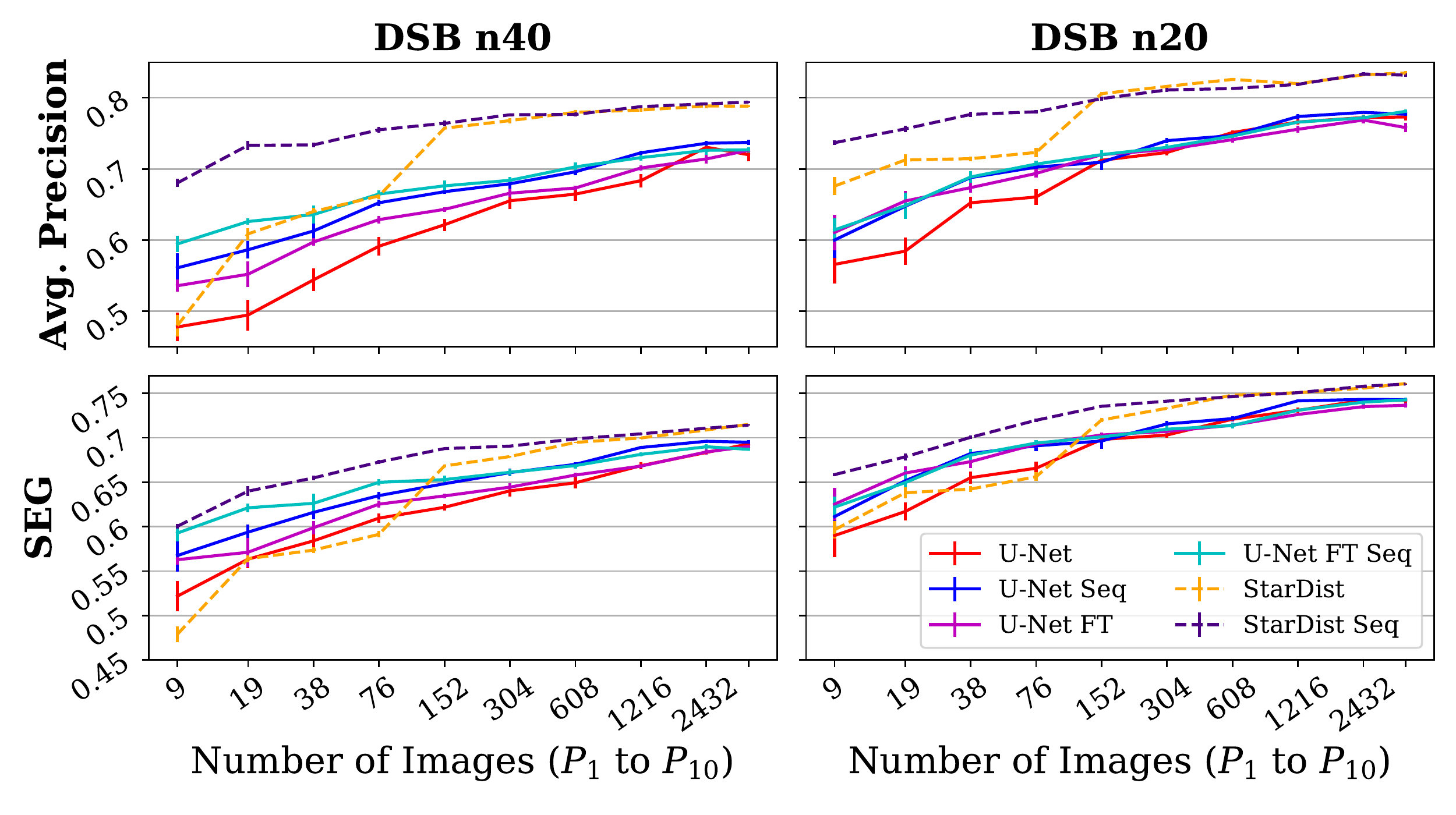}
}
\caption{Results for noise level n40 and n20 on DSB data. Sequential is abbreviated as Seq and Finetune is abbreviated as FT.
It can be seen that our proposed training schemes consistently outperform the respective baseline, mainly when only limited segmentation GT is available. }
\label{fig:DSBSEGAP}
\end{figure}
}

\newcommand\figBBBCSEGAP{
\begin{figure}[t]
\centerline{
\includegraphics[width=\linewidth]{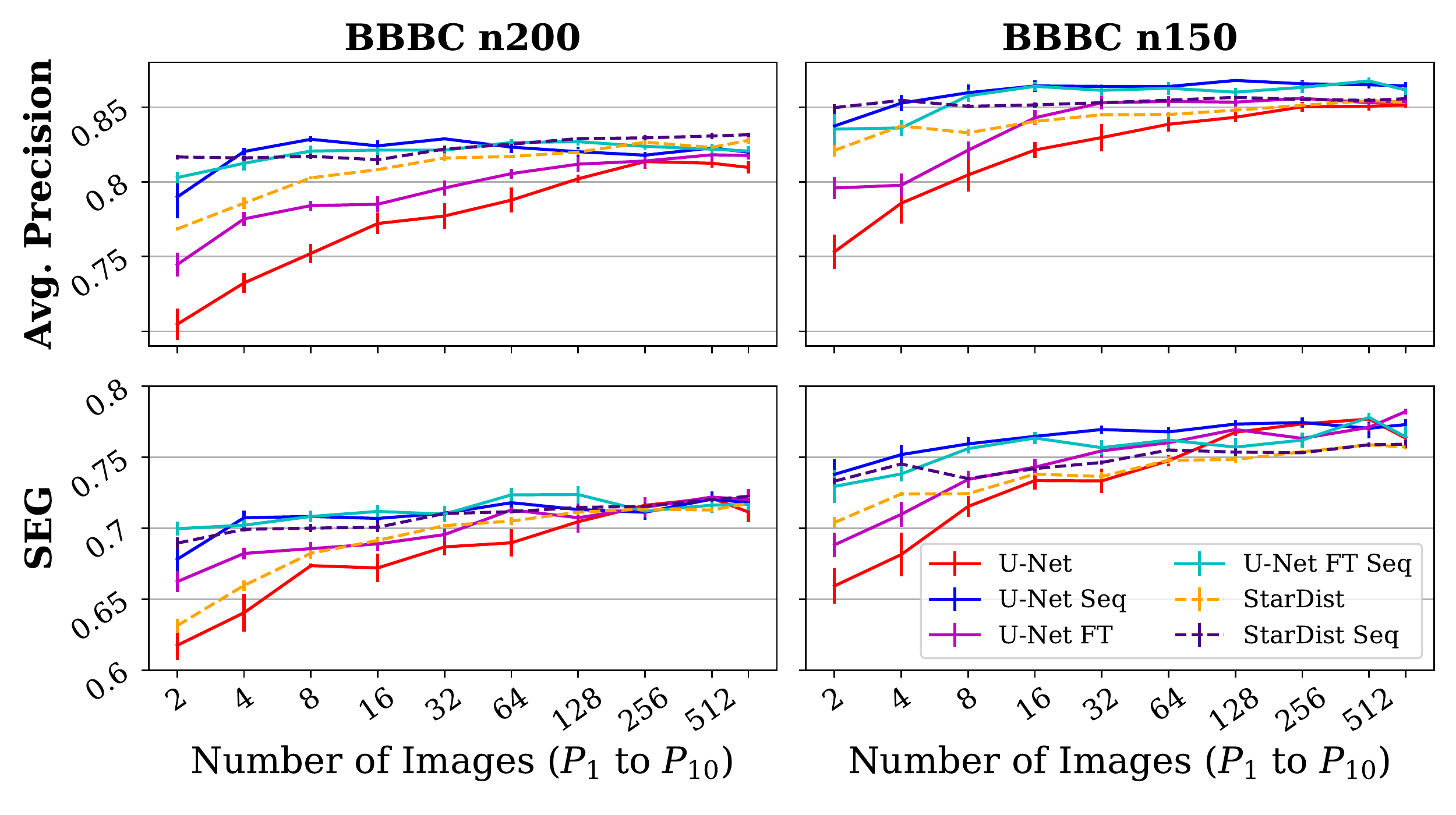}
}
\caption{Results for noise level n200 and n150 on BBBC data. The abbreviations are the same as in Fig.~\ref{fig:DSBSEGAP}.
Again, all proposed training schemes outperform their baselines.
Here our proposed sequential U-Net schemes even outperform StarDist and StarDist Sequential.}
\label{fig:BBBCSEGAP}
\end{figure}
}

\newcommand\figQualitative{
\begin{figure*}[tb]
\centerline{
\includegraphics[trim=0pt 510pt 0pt 100pt,scale = 0.4, clip=true,width=\linewidth]{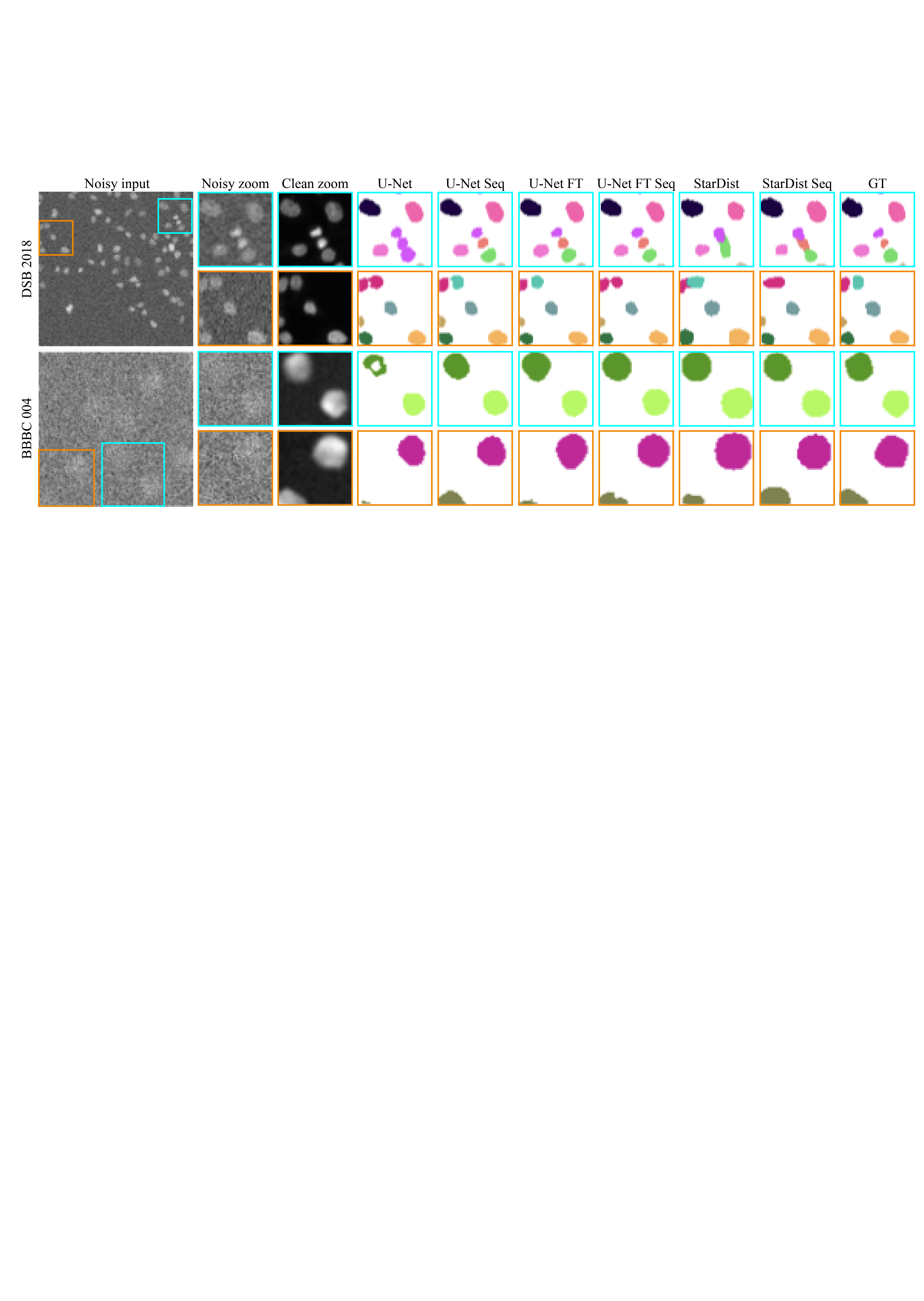}
}
\caption{Visual comparison of segmentation results with baseline methods and proposed training schemes for DSB n40 $P_1$ and BBBC n200 $P_1$. 
From left to right we show first one noisy input image, then the two insets, respective noise-free data, and the various segmentation results with each object shown in a distinct color. Sequential is abbreviated as Seq and Finetune is abbreviated as FT.
In line with the overall performance on the full body of data, also in the examples we show our proposed methods outperform the quality achieved by the baselines.
}
\label{fig:Qualitative}
\end{figure*}
}

\newcommand\tabDsbBbcSubset{
\begin{table*}[h]
\begin{tabular}{ p{2.9cm}||p{1cm} p{1cm}p{1cm}p{1cm}p{1cm}p{1cm}p{1cm}p{1cm}p{1cm}p{1cm}p{1cm}  }
  \hline
  \multicolumn{11}{c}{DSB 2018 n40} \\
   \hline
    Scheme & $P_1$ & $P_2$ & $P_3$ & $P_4$ & $P_5$ & $P_6$ & $P_7$ & $P_8$ & $P_9$ & $P_{10}$ \\
 \hline
 U-Net  & 0.4777, & 0.4944, & 0.5439, & 0.5912, & 0.6214, & 0.6551, & 0.6645, & 0.6834, & 0.7304, & 0.7199,\\ 
 & \textit{0.5218} & \textit{0.5634} & \textit{0.5840} & \textit{0.6095} & \textit{0.6217} & \textit{0.6403} & \textit{0.6493} & \textit{0.6685} & \textit{0.6835} & \textit{0.6929}\\ 
  \hdashline
 U-Net Sequential & 0.5608, & 0.5862, & 0.6127, & 0.6523, & 0.6679, & 0.6791, & 0.6958, & \textbf{0.7226}, & \textbf{0.7360}, & \textbf{0.7373},\\ 
 & \textit{0.5675} & \textit{0.5938} & \textit{0.6160} & \textit{0.6349} & \textit{0.6483} & \textit{0.6608} & \textbf{\textit{0.6700}} & \textbf{\textit{0.6890}} & \textbf{\textit{0.6960}} & \textbf{\textit{0.6950}}\\ 
 \hdashline
 U-Net Finetune & 0.5357, & 0.5518, & 0.5971, & 0.6286, & 0.6430, & 0.6658, & 0.6731, & 0.7013, & 0.7140, & 0.7261,\\
 & \textit{0.5628} & \textit{0.5711} & \textit{0.5987} & \textit{0.6253} & \textit{0.6346} & \textit{0.6444} & \textit{0.6580} & \textit{0.6681} & \textit{0.6840} & \textit{0.6901}\\
 \hdashline
 U-Net Finetune Seq. & \textbf{0.5944}, & \textbf{0.6259}, & \textbf{0.6357}, & \textbf{0.6646}, & \textbf{0.6761}, & \textbf{0.6839}, & \textbf{0.7028}, & 0.7158, & 0.7261, & 0.7267,\\
 & \textbf{\textit{0.5927}} & \textbf{\textit{0.6212}} & \textbf{\textit{0.6262}} & \textbf{\textit{0.6499}} & \textbf{\textit{0.6529}} & \textbf{\textit{0.6611}} & \textit{0.6686} & \textit{0.6813} & \textit{0.6898} & \textit{0.6870}\\
 \hline
 StarDist & 0.4796, & 0.6085, & 0.6400, & 0.6620, & 0.7572, & 0.7679, & \textbf{0.7795}, & 0.7827, & 0.7884, & 0.7883,\\
& 0.4789 & \textit{0.5639} & \textit{0.5735} & \textit{0.5913} & \textit{0.6683} & \textit{0.6788} & \textit{0.6948} & \textit{0.6997} & \textit{0.7087} & \textbf{\textit{0.7150}}\\
\hdashline
 StarDist Sequential & \textbf{0.6802,} & \textbf{0.7331,} & \textbf{0.7337,} & \textbf{0.7549,} & \textbf{0.7640,} & \textbf{0.7761,} & 0.7766, & \textbf{0.7876,} & \textbf{0.7914,} & \textbf{0.7939,}\\
 & \textbf{\textit{0.6004}} & \textbf{\textit{0.6399}} & \textbf{\textit{0.6548}} & \textbf{\textit{0.6727}} & \textbf{\textit{0.6877}} & \textbf{\textit{0.6906}} & \textbf{\textit{0.6987}} & \textbf{\textit{0.7044}} & \textbf{\textit{0.7107}} & \textit{0.7141}\\
 \hline
 \multicolumn{11}{c}{BBBC 004 n200} \\
   \hline
    Scheme & $P_1$ & $P_2$ & $P_3$ & $P_4$ & $P_5$ & $P_6$ & $P_7$ & $P_8$ & $P_9$ & $P_{10}$ \\
 \hline
 U-Net  & 0.7046, & 0.7323, & 0.7520, & 0.7721, & 0.7771, & 0.7877, & 0.8020, & 0.8135, & 0.8124, & 0.8096,\\ 
 & \textit{0.6175} & \textit{0.6405} & \textit{0.6736} & \textit{0.6720} & \textit{0.6868} & \textit{0.6897} & \textit{0.7046} & \textbf{\textit{0.7161}} & \textit{0.7211} & \textit{0.7114}\\
 \hdashline
 U-Net Sequential & 0.7898, & \textbf{0.8202}, & \textbf{0.8284}, & \textbf{0.8240}, & \textbf{0.8287}, & 0.8232, & 0.8201, & 0.8178, & \textbf{0.8226}, & 0.8198,\\ 
 & \textit{0.6781} & \textbf{\textit{0.7074}} & \textbf{\textit{0.7083}} & \textit{0.7069} & \textbf{\textit{0.7106}} & \textit{0.7179} & \textit{0.7132} & \textit{0.7112} & \textit{0.7205} & \textit{0.7179}\\ 
 \hdashline
 U-Net Finetune & 0.7446, & 0.7752, & 0.7840, & 0.7850, & 0.7959, & 0.8055, & 0.8118, & 0.8139, & 0.8180, & 0.8176,\\
 & \textit{0.6624} & \textit{0.6822} & \textit{0.6856} & \textit{0.6889} & \textit{0.6955} & \textit{0.7129} & \textit{0.7073} & \textit{0.7148} & \textbf{\textit{0.7219}} & \textbf{\textit{0.7201}}\\
 \hdashline
 U-Net Finetune Seq. & \textbf{0.8028}, & 0.8124, & 0.8206, & 0.8212, & 0.8212, & \textbf{0.8263}, & \textbf{0.8268}, & \textbf{0.8237}, & 0.8218, & \textbf{0.8206},\\
 & \textbf{\textit{0.6996}} & \textit{0.7021} & \textit{0.7082} & \textbf{\textit{0.7118}} & \textit{0.7099} & \textbf{\textit{0.7235}} & \textbf{\textit{0.7237}} & \textit{0.7120} & \textit{0.7162} & \textit{0.7168}\\
 \hline
 StarDist & 0.7684, & 0.7857, & 0.8026, & 0.8081, & 0.8159, & 0.8169, & 0.8199, & 0.8264, & 0.8230, & 0.8277,\\
 & \textit{0.6313} & \textit{0.6597} & \textit{0.6823} & \textit{0.6914} & \textit{0.7018} & \textit{0.7050} & \textit{0.7113} & \textit{0.7135} & \textit{0.7128} & \textit{0.7174}\\
 \hdashline
 StarDist Sequential & \textbf{0.8166}, & \textbf{0.8159}, & \textbf{0.8170}, & \textbf{0.8147}, & \textbf{0.8219}, & \textbf{0.8251}, & \textbf{0.8289}, & \textbf{0.8294}, & \textbf{0.8306}, & \textbf{0.8314},\\
 & \textbf{\textit{0.6895}} & \textbf{\textit{0.6992}} & \textbf{\textit{0.7001}} & \textbf{\textit{0.7007}} & \textbf{\textit{0.7103}} & \textbf{\textit{0.7116}} & \textbf{\textit{0.7146}} & \textbf{\textit{0.7153}} & \textbf{\textit{0.7202}} & \textbf{\textit{0.7224}} \\
\hline
\end{tabular}
\label{TableDsbBbcSubset}
\caption{Mean performance in terms of average precision (AP) and SEG (in italic) for DSB n40 (8 repetitions) and for BBBC n200 (5 repetitions). Bold number indicate the best performing scheme for a given fraction of segmentation GT ($P_i$).
See the main text for further details.}
\vspace{2mm}
\end{table*}
}

\begin{document}
%
\maketitle
\begin{abstract}
Deep learning (DL) has arguably emerged as the method of choice for the detection and segmentation of biological structures in microscopy images. 
However, DL typically needs copious amounts of annotated training data that is for biomedical projects typically not available and excessively expensive to generate. 
Additionally, tasks become harder in the presence of noise, requiring even more high-quality training data.
Hence, we propose to use denoising networks to improve the performance of other DL-based image segmentation methods. 
More specifically, we present ideas on how state-of-the-art self-supervised CARE networks can improve cell/nuclei segmentation in microscopy data. 
Using two state-of-the-art baseline methods, U-Net and StarDist, we show that our ideas consistently improve the quality of resulting segmentations, especially when only limited training data for noisy micrographs are available.
\end{abstract}
%
%
%
%
%
\section{Introduction}
\label{sec:intro}
Modern microscopy techniques enable us to capture biological processes at high spatial and temporal resolution.
Today, the total amount of acquired image data can be so vast, analyzing it poses a tremendous challenge. 
In many cases, deep learning (DL) based analysis methods~\cite{litjens2017survey,falk2019u,rutter2019convolutional,schmidt2018} are leading the way to address this problem.
Still, even common tasks, such as detection or segmentation, typically require human curation to fix remaining errors.

Arguably the two main causes of weak detection and segmentation performance are 
$(i)$~too little training data, and
$(ii)$~input images acquired at low signal-to-noise ratios (SNR).
In order to make most out of the available training data, augmentation~\cite{simard2003best} and transfer learning~\cite{doersch2015unsupervised, zamir2018taskonomy} are often used.
While data augmentation uses transformed copies of the training data to gain better training performance, transfer learning employs networks pretrained on similar tasks and/or data and finetunes them for the task/data at hand.
To address the low SNR, a number of powerful content-aware restoration and denoising methods have recently been developed~\cite{zhang2017beyond, weigert2018content,lehtinen2018noise2noise}.

Among them are self-supervised
methods~\cite{krull2019noise2void,batson2019noise2self,krull2019probabilistic}, which do not require annotated training data, and can be directly trained on the raw data to be denoised.

\figTeaser
In this work, we investigate various ways, in which self-supervised denoising can enable cell/nuclei segmentation, even in the presence of extreme levels of noise and limited training data. 
We explore the efficacy of denoising as a preprocessing step, as part of a transfer learning schema, as well as, in a combination of the two.

We conducted all experiments with two popular DL-based segmentation methods:
a standard U-Net~\cite{ronneberger2015u, falk2019u} and the more sophisticated StarDist~\cite{schmidt2018}.
While we find that self-supervised denoising generally improves segmentation results, especially when noise is abundant and training data limitied, we provide detailed results, comparing all approaches for various amounts of training data, noise levels, and types of data.
All datasets, results, and code can be found at \small{\url{github.com/juglab/VoidSeg}}.

\section{Methods and Experiments}
\label{sec:methods}
\vspace{\chapterspacing}

\figDSBSEGAP

As sketched in Fig.~\ref{fig:Teaser}, we propose three ways to improve segmentations and compare our results to two baseline methods, namely 
$(i)$~a standard U-Net~\cite{ronneberger2015u} for 3-class pixel classification, and
$(ii)$~StarDist~\cite{schmidt2018}, designed to learn and utilize a star convex shape prior.
These baselines are chosen based on popularity and because they follow rather different segmentation paradigms.
The following setups are the ones we propose.

\compactsubsub{Sequential (Figure~\ref{fig:Teaser}b)} {
Here, two networks are employed. 
The first network is a Noise2Void (N2V) network~\cite{krull2019noise2void}, trained to denoise the full body of available image data. 
The second network, which henceforth receives the denoised N2V output, is then either a U-Net or StarDist network, trained on all or parts of the available segmentation labels (GT). 
Note that all weights of the N2V network remain constant during training the segmentation network.
}

\compactsubsub{Finetune (Figure~\ref{fig:Teaser}c)}{
In contrast to the sequential setup, here we retrain the N2V network for segmentation.
Since StarDist does not use the exact same network architecture as N2V, this approach only applies to the U-Net baseline. 
}

\compactsubsub{Finetune Sequential (Figure~\ref{fig:Teaser}d)}{
Very similar to the sequential setup, also here we first train a N2V denoising network. 
In contrast to before, the segmentation network is initialized by a copy of the trained N2V network and then finetuned for segmentation.
Also here, the weights of the first network stay unchanged during the training of the segmentation network.
}


\vspace{2mm}
\noindent Next we describe the detailed setup of the N2V, Segmentation \mbox{U-Net}, and StarDist networks.

\compactsubsub{N2V Denoising Network}{
We use the Noise2Void setup as described in~\cite{krull2019noise2void}. 
Conveniently, N2V is just a default U-Net with a modified loss for denoising, allowing us to design a single network that can later be used for N2V training as well as for the U-Net segmentation baseline.
We use $32$ initial feature maps with batch norm and a batch size of $128$ and employ $3 \times 3$ convolution kernels. 
For all experiments we choose the depth of the U-Net as described below.
}

\compactsubsub{U-Net Segmentation Network}{
We created a U-Net capable of performing either 3-class pixel classification (foreground, border, background) \cite{chen2016dcan, guerrero2018multiclass} or N2V denoising.
Hence, the U-Net we use has four output channels, one for each pixel class, and one to regress denoised pixel intensities.
Note that, during pixel classification, we give extra emphasis to the border class, by weighing it five times higher in the used loss as suggested in~\cite{schmidt2018}.
Again we use $32$ feature maps, batch size of $128$, and $3 \times 3$ kernels. 
For all experiments, the depth of the U-Net is chosen to saturate segmentation performance (making the network deeper would not lead to improved results).
Hence, results below are not limited by the capacity of network.
All networks are trained with a standard learning rate scheduler as used in \cite{krull2019noise2void}.
We use an initial learning rate of $0.0004$ and a batch size of $128$ with batch normalization.
Training is done for $200$ epochs, each consisting of $400$ steps. 
Training data is augmented 8 fold by flips and 90 degree rotations.
}

\compactsubsub{StarDist Segmentation Network}{
Number of feature maps, batch size, convolution kernels, network depth, learning rate, number of training epochs, and step size per epoch used for StarDist are set as described above. Again, the training data is augmented 8 fold by flips and 90 degree rotations.
However, StarDist uses $33$ output channels that are trained as described in~\cite{schmidt2018}.
}

\section{Data and Evaluation Metrics}
\label{sec:datasets}
\vspace{\chapterspacing}

\figBBBCSEGAP

\tabDsbBbcSubset

In this work we use publicly available data, which we randomly split into training and test sets (see following subsections for details). 
We further split the training data into ${P_1\subset P_2\subset\ldots\subset P_{10}}$, ten stacked subsets we will use to evaluate our methods in data-limited training regimes.
Additionally, we corrupt the raw microscopy data with pixel independent, identically distributed Gaussian noise.
Sample images for all datasets are shown in Fig.~\ref{fig:Qualitative}.

\compactsubsub{DSB 2018 Data}{
From the Kaggle 2018 Data Science Bowl challenge, we take the same subset of data as has been used in~\cite{schmidt2018}, showing a diverse collection of cell nuclei imaged by various fluorescence microscopes. 
We extracted $4470$ image patches of size $128\times 128$ from the training set.  
For this data, manually generated segmentation GT is available. 
Training subsets $P_1$ through $P_{10}$ consist of $10,19,38,76,152,304,608,1216,2432,3800$ randomly chosen image patches, respectively.
The remaining $670$ patches constitute the validation set while the test set has $50$ additional images of different sizes.
Additional noise is added with mean $0$ and standard deviations $10$, $20$, and $40$ to training, validation and test data. 
We refer to the modified datasets as n10, n20, and n40, respectively.
}

\compactsubsub{BBBC 004 Data}{
This data is available from the Broad Bioimage Benchmark Collection and consists of synthetic nuclei images. 
Since the data is synthetic, perfect GT labels are available by construction. 
Here we use only the images having non-touching nuclei.
We extracted $880$ image patches (of size $128\times 128$) from the training set.
Training subsets $P_1$ through $P_{10}$ consist of $2,4,7,15, 30,60,120,239,479,748$ image patches, respectively while the validation set consists of remaining 132 patches. The test set consists of additional $220$ patches.
Additional noise is added with mean $0$ and standard deviations $150$ and $200$ to training and test data. 
Following the naming convention from above, we refer to this data as n150 and n200.
}

All experiments we conduct are evaluated in terms of Average Precision (AP)~\cite{everingham2010pascal} and SEG~\cite{ulman2017objective}.
The SEG measure is based on the Jaccard similarity index ($J$), computed for matching objects $S$ and $R$, and is given by $J(S,R)=(|R \cap S|)/(|R \cup S|)$.
A ground truth object $R$ and a segmented object $S$ are considered to be matching if and only if at least $50\%$ of the pixels of $R$ are overlapped by pixels in $S$. 
Average Precision, in contrast, counts the ratio of true positives to the sum of true positives, false positives, and false negatives.
All AP and SEG values we report here are obtained by finding the threshold that maximizes AP.
For the U-Net this threshold is used to cut the foreground probability maps into discrete image regions.
For StarDist the threshold controls the non-maxima suppression step~\cite{schmidt2018}.

\section{Results}
\label{sec:results}
\vspace{\chapterspacing}

\figQualitative

We investigated all setups described above, on all noise levels, using all $10$ subsets of training data $P_i$, making a total of $60$ experimental setups. 
Each experiment on the DSB data was repeated $8$ times while all experiments on the BBBC data were repeated $5$ times, allowing us to report mean performance and standard error for selected noise levels in Figures~\ref{fig:DSBSEGAP} and \ref{fig:BBBCSEGAP}.
Additionally we show results for DSB data at noise level n40 and results for BBBC data at n200 in Table~\ref{TableDsbBbcSubset}. 
A complete set of figures and tables, for all conducted experiments, can be found online at \small{\url{github.com/juglab/VoidSeg/wiki}}.

Looking at all results it can be observed that all our proposed schemes outperform their respective baseline when the amount of available training data is limited. The Finetune Sequential scheme is typically performing best among all U-Net based pixel classification pipelines. 
StarDist, in itself a more powerful method, is indeed the better performing baseline.
As before, the proposed StarDist Sequential scheme clearly outperforms its baseline method when fewer training images are available.
Note that even if ample training data is provided, all our proposed training schemes perform at least on par with their baselines.
It is important to be reminded that improved results using sequential training schemes are not due to limiting network sizes -- we have tested various network sizes for both baseline methods and have settled for the best performing configuration we could find.

To our surprise, on the BBBC data, both sequential U-Net schemes outperform StarDist and StarDist Sequential for the n150 and n200 noise levels, despite the StarDist baseline consistently and significantly outperforming the U-Net baseline.

A visual comparison of segmentation results with all the methods trained on the training subset $P_1$ is given in Fig.~\ref{fig:Qualitative}.
For the DSB data we show insets that exemplify the often occurring problem of merging segments (bad for AP), while the shown BBBC insets show variations in segmented areas (bad for SEG). These segmentation mistakes are particularly exemplified for baseline schemes whereas sequential schemes for both U-Net and StarDist seem to yield better quality segmentation, in general.

\section{Discussion}
\label{sec:discussion}
\vspace{\chapterspacing}
It is known that there is an overlap between denoising and segmentation tasks~\cite{zamir2018taskonomy}.
In this work we investigated how disentangling the two can be exploited in practice, when noisy data is abundant, but annotations are rare -- a situation that is virtually ubiquitously true in biomedical applications.

In these situations, all our proposed schemes show above baseline performance.
Among all conducted experiments, sequential training schemes generally lead to the best results.
Since this is not only true for the simple U-Net baseline, but also for StarDist, it stands to reason that similar observations would also hold for other DL based approaches and tasks. 
Here we do not test other segmentation approaches, but we believe that N2V, or other self-supervised denoising methods~\cite{krull2019probabilistic,batson2019noise2self}, can serve as universal preprocessing blocks for networks solving any given super-task in which denoising is a helpful sub-task. 

These denoising blocks can benefit from the whole body of available noisy data, without relying on annotated GT labels required for the super-task.
Hence, finding sensible training schedules to train such larger, modular networks is a promising direction of research.

In summary, we show that commonly used networks for image segmentation can likely be boosted in performance by combining them in various ways with unsupervised denoising modules. 
Our work offers simple recipes for improving DL based segmentation results. 
Since this is increasingly true at lower signal-to-noise regimes and when segmentation GT is limited, direct benefits for the biomedical imaging community will be inevitable.

\section{Acknowledgements}
\label{sec:acks}
\vspace{\chapterspacing}
The authors would like to acknowledge the Scientific Computing Facility at MPI-CBG and HPC Cluster at the Center for Information Services and High Performance Computing (ZIH) at TU Dresden for giving us access to their HPC cluster. We also thank Uwe Schmidt, Martin Weigert and Vladimir Ulman from MPI-CBG for helpful discussions. This work was supported by the German Federal Ministry of Research and Education (BMBF) under the code 01IS18026C (ScaDS2).


\bibliographystyle{IEEEbib}
\bibliography{refs}

\end{document}